\documentclass{article}

\usepackage[font=footnotesize]{caption}
\usepackage[utf8]{inputenc}
\usepackage[T1]{fontenc}
\usepackage{lmodern}
\usepackage{xcolor}
\usepackage{graphicx}
\usepackage{authblk}
\usepackage{url}
\usepackage{subcaption}
\usepackage{indentfirst}
\usepackage{amsmath}
\usepackage[title]{appendix}
\usepackage{hyperref}
\usepackage[left=2cm,right=2cm,top=2.5cm,bottom=2.5cm]{geometry}

\DeclareMathOperator{\erf}{erf}

\title{The Jaccard Similarity Mean}
\author{Gonzalo Travieso}

\author{Luciano da F. Costa}

\affil{São Carlos Institute of Physics, University of São Paulo, São Carlos, SP, Brazil}

\begin{document}

\maketitle

\begin{abstract}
The arithmetic mean plays a central role in science and technology, being directly related to the concepts of statistical expectance and centrality.  Yet, it is highly susceptible to the presence of outliers or biased interference in the original dataset to which it is applied.  Described recently, the concept of similarity means has been preliminary found to have marked robustness to those same effects, especially when adopting the Jaccard similarity index.  The present work is aimed at investigating further the properties of similarity means, especially regarding their range, translating and scaling properties, sensitivity and robustness to outliers.  Several interesting contributions are reported, including an effective algorithm for obtaining the similarity mean, the analytic and experimental identification of a number of properties, as well as the confirmation of the potential stability of the similarity mean to the presence of outliers. The present work also describes an application case-example in which the Jaccard similarity is successfully employed to study cycles of sunspots, with interesting results.
\end{abstract}

\section{Introduction}

The concept of \emph{mean}, while not formally definable in mathematical terms (e.g.,~\cite{Bertsekas,Kreyszig,DeGroot}), holds significant importance in science and technology due to its association with \emph{summarizing} a set of quantitative values by capturing their central tendency, retaining the same unit as the original data.  In a particular sense, the concept of means can be understood as an approach to modeling (e.g.~\cite{CostaModeling}) a set of observed measurements in terms of a single representative number.

The \emph{arithmetic mean} constitutes, arguably, the prototypical case among the several types of means.  It not only relates directly to the concept of \emph{expectance} of a random variable in the areas of probability and statistics, but also corresponds to the center of mass when approached from physics and mechanics (e.g.~\cite{goldstein2002}).

While the arithmetic, geometric and harmonic means have been studied at least from the time of the Pythagoreans, being known as \emph{Pythagorean}, there is potentially no limit to the number of alternative types of means, which also encompass the weighted arithmetic, Lehmer (e.g.~\cite{Bullen}) and root mean square approaches.  At the same time, all these means share at least the fact of being functionals (mapping a set of values into a scalar) and having the same physical unit as the original values to be summarized~\cite{CostaSimMean,CostaRobust}.

Another property that is often expected from a mean concerns its potential \emph{stability} to perturbations of the dataset being summarized.  Such perturbations may include acquisition noise, round-off, systematic and/or intermittent errors, interference between groups, as well as possible presence of outlier values.

Despite its central importance, the arithmetic mean proves surprisingly sensitive to outliers~\cite{Recipes}. To address this limitation, alternative approaches have been proposed, such as the concept of similarity mean, which has shown promising robustness to outliers and data perturbations~\cite{CostaSimMean}. Similarity means are derived from respective similarity indices, which quantify the degree of similarity between values (e.g.~\cite{sorensen1948,wolda1981,Lieve1989,kabir2017,steinley2021}). For example, the Jaccard similarity mean is calculated by considering the Jaccard similarity index~\cite{jaccard1901,jac:wiki}. In more detail, given a set of values, its similarity mean corresponds to the value that maximizes the relevant similarity index with respect to those values.

The \emph{main motivation} for considering and employing similarity means, including those based on the Jaccard and coincidence (e.g.~\cite{CostaJaccard,CostaSimilarity,CostaMNeurons,CostaCCompl}) similarity indices, concerns the possibility of quantifying the centrality of a dataset in terms of the concept of \emph{overall similarity} among data elements instead of adopting other criteria including the physical principle of center of mass (centroid) or statistical moments (e.g.~\cite{johnsonwhichern}).  Indeed, the concept of similarity means can be understood as the value which is \emph{most similar} to the whole set of elements in a dataset.  In addition to allowing this interesting possibility, the consideration of similarity means can also lead to intrinsic properties inherited from the similarity indices.  In the case of the Jaccard and coincidence indices, it has been shown (e.g.~\cite{CostaMNeurons,CostaCCompl}) that enhanced sensitivity in the quantification of the similarity between data elements can be achieved, allowing more strict comparisons which can, if necessary, be controlled in terms of an exponent $D$.  In addition, these indices have been found to be particularly robust to the presence of noise in the data measurements, as well as to respective outliers.  Yet another potential advantage of those similarity indices concerns the fact that they take non-dimensional values that are inherently normalized between 0 and 1.

A first relevant aspect pertaining the study of a similarity mean concerns the characterization of the possible values it can take.  In particular, it becomes interesting to know how these values relate to the original set of values being summarized by the mean.  Yet another property that is especially important to a given type of mean concerns the fact whether it is unique or not (degree of degeneracy) among the considered domain.  In addition, it would be interesting to know how a similarity mean changes when the original data is translated or scaled.

The present work aims at studying the Jaccard similarity mean from the perspective of the above motivated issues and properties.  In particular, we aim at quantifying analytically and experimentally the robustness of this mean to the presence of a group of outlier values.

The work starts by presenting the basic concepts, including multiset theory, the Jaccard similarity index, as well as the concept of similarity means.  Then, it is shown that the Jaccard mean is necessarily contained in the original set of values.  An effective algorithm for estimating the Jaccard mean is presented next, which is then followed by its analytic calculation given a discrete or continuous probability distribution.  As examples we use the uniform, exponential, truncated normal, and power law distributions.  We then define a coefficient that quantifies variability in the data based on the Jaccard mean. The relationship between the Jaccard mean and the median, as well as its modifications in presence of data translation and scaling are presented subsequently.   Then, the robustness of the Jaccard mean to outliers is addressed, including the case when the data is translated.  A methodology for identifying the skewness and possible outliers in datasets is presented in the following, including a complete application example to the analysis of sunspots.

\section{Basic Concepts}  \label{sec:basic}

This section provides a review of the basic concepts adopted in the present work, including principles of multisets, the Jaccard simialarity index, and the concept of similarity mean.

\subsection{Multisets}

\emph{Sets} (e.g.~\cite{Kreyszig}) are mathematical structures which have played a central role in the physical sciences from its very beginnings.  Its ubiquity results from the ability of a set to serve as a model of several situations in which it is necessary to identify, in any order, the types of objects or entities in a given collection.  Another important property of a set is that each of its \emph{elements} can appear only once.

\emph{Multisets} (e.g.~\cite{CostaMSet}) constitute an extension of the concept of set in the sense of allowing an \emph{element to appear multiple times}, which is expressed in terms of its respective \emph{multiplicity}.  Thus, multisets can be represented as sets of tuples $(x_i,m_i)$ corresponding to each of the elements $x_i$, associated to its multiplicity $m_i$.   The set of all possible elements in a multiset can be expressed in terms of the \emph{support} of that multiset.  Multisets can also be represented as sets with repeated elements.

As an example, consider the two following multisets:
\begin{align}
   &A = \left\{  a, a, a, b, c, c \right\} =
   \left\{(a, 3); (b, 1); (c, 2)  \right\}   \\ 
   &B = \left\{  a, c, c, c, d \right\} =
   \left\{(a, 1); (c, 3), (d, 1)  \right\} 
\end{align}
with supports $S_A = \left\{a, b, c \right\}$ and $S_B = \left\{a, c, d \right\}$, respectively.

Though originally defined for non-negative integer multiplicities, multisets can be readily extended~\cite{CostaJaccard,CostaSimilarity,CostaMNeurons} to virtually any mathematical structure, including vectors, functions, matrices, graphs, etc.

In the case of two multisets $A$ and $B$ consisting of real-valued vectors with non-negative entries, their \emph{union} can be expressed in terms of the multiset $C$ with support consisting of the union of the supports of $A$ and $B$, and taking the multiplicity of each of these elements as corresponding to the respective maximum between their original multiplicities in $A$ and $B$.   As an example, in the case of the two multisets above, we have:
\begin{align}
   &C = A \cup B = \left\{  a, a, a, b, c, c, c, d \right\} =
   \left\{(a, 3); (b, 1); (c, 3); (d,1)  \right\} 
\end{align}
with $S_C = \left\{ a, b, c, d \right\}$.

The \emph{intersection} between $A$ and $B$ is similarly defined, but taking the minimum multiplicity instead of the maximum.  As  an example, in the case of the above multisets, it follows that:
\begin{align}
   &D = A \cap B = \left\{  a, c, c \right\} =
   \left\{(a, 1); (c, 2) \right\} 
\end{align}
with $S_D = \left\{ a, c \right\}$.

\begin{figure}
\centering
\begin{subfigure}{0.4\textwidth}
    \includegraphics[width=\textwidth]{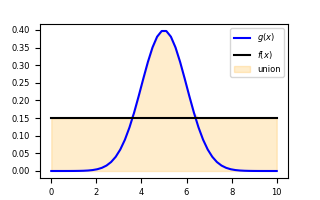}
    \caption{}
    \label{fig:union}
\end{subfigure}
\begin{subfigure}{0.4\textwidth}
    \includegraphics[width=\textwidth]{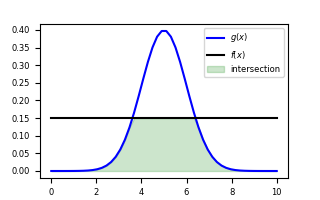}
    \caption{}
    \label{fig:intersection}
\end{subfigure}
    \caption{Illustration of the generalization of the union and intersection operations for (non-negative) functions: Union (a) and intersection (b) of $f(x)=0.15$ with $g(x)=\frac{1}{\sqrt{2\pi}}e^{-\frac{1}{2}(x-5)^2}$ in the interval $[0,10]$.}
    \label{fig:setoperations}
 \end{figure}

The multiset union and intersection operations between functions can be conveniently visualized.   For instance, Figure~\ref{fig:setoperations} shows the union and intersection between a constant function $f(x) > 0$ and a Gaussian function $g(x) > 0$.  

As illustrated in that figure, the multiset union corresponds to the union of the areas of the two functions, being therefore implemented by the maximum operation at each of the values of $x$.  The multiset intersection of the areas corresponds to the minimum between the functions.

\subsection{Jaccard Similarity}\label{sec:definition}

Let $\pmb{x}$ and $\pmb{y}$ be two non-zero $N \times 1$ dimensional vectors of non-negative values.  The \emph{Jaccard similarity index} between these two vectors can be expressed as (e.g.~\cite{CostaJaccard,CostaSimilarity,CostaCCompl}):
\begin{align}
  \mathcal{J}(\pmb{x},\pmb{y}) = \frac{\sum_{i=1}^N \min \left\{ x_i,y_i \right\} }  {\sum_{i=1}^N \max \left\{ x_i,y_i \right\}}.
\end{align}
with $0 <   \mathcal{J}(\pmb{x},\pmb{y}) \leq 1$ and $\mathcal{J}(\pmb{x},\pmb{y}) =   \mathcal{J}(\pmb{y},\pmb{x})$.  

In the expression above, the $\min$ and $\max$  correspond to the multiset (e.g.~\cite{CostaMSet}) operations with non-negative multiplicity respective to the set union and intersection operations from classic set theory.  In addition, observe that the Jaccard operator is not \emph{bilinear}, which contributes to its  more intricate properties as compared to, e.g., the scalar product (e.g.~\cite{CostaScalar}).

We can define \cite{CostaSimMean} the Jaccard similarity of a scalar value $x$ with the vector $\pmb{x}$ as
\begin{align}\label{eq:jackx}
  \mathcal{J}(x,\pmb{x}) = \frac{\sum_{i=1}^N \min \left\{ x, x_i \right\} }  {\sum_{i=1}^N \max \left\{ x, x_i \right\}}.
\end{align}

Observe that the order of the elements of $\pmb{x}$ does not matter, so that the above definition holds also for a non-empty set $X$ of non-negative values.

Given a random variable $X$ with a discrete set of possible values $x_i$ for $i=1,2\ldots$, each $x_i$ with an associated probability $p_X(x_i)$, we can define the Jaccard similarity of a value $x$ with the random variable $X$ as
\begin{align}
  \mathcal{J}_X(x) = \frac{\sum_{i=1}^\infty \min \left\{ x, x_i \right\} p_X(x_i)}  {\sum_{i=1}^\infty \max \left\{ x, x_i \right\} p_X(x_i)}.\label{eq:jaccdist}
\end{align}
Notice that this is equivalent to Eq.~\eqref{eq:jackx} if we consider the $x_i$ here as the unique values in the vector \pmb{x} used in Eq.~\eqref{eq:jackx} and the $p_i$ here as $n_i/N$, where $n_i$ is the multiplicity of the value $x_i$ in the vector. 

If we have a continuous random variable $X$ with a given probability density function (PDF) $\rho_{X}(x)$ we can define the Jaccard similarity of $x$ with $X$ as
\begin{align}
  \mathcal{J}_X(x) & = \frac{\int_0^\infty \min\{x,\xi\}\rho_X(\xi)\,d\xi}{\int_0^\infty \max\{x,\xi\}\rho_X(\xi)\,d\xi}.\label{eq:jcontdef}
\end{align}
Notice that this is a generalization of Eq.~\eqref{eq:jaccdist} if we use 
$$\rho_X(x) = \sum_{i=1}^\infty p_X(x_i)\delta(x - x_i),$$
with $\delta(x)$ the Dirac delta.

\subsection{Similarity Means}

The concept of similarity mean has been recently suggested~\cite{CostaSimMean,CostaRobust} as follows.  Given a set of $N$ values $S$, which may correspond to a set of observations of a given random variable $X$, as well as a specific type of similarity index (e.g.~Jaccard), the respective similarity mean corresponds to the value of $x$ comprised in the support of $X$ that yields the maximum similarity index.  Observe, however, that as defined the similarity mean does not necessarily have to correspond to one of the elements of the set $S$.

The main motivation underlying the concept of similarity mean consists in the idea of obtaining a value $x$ that is as close as possible to the \emph{whole set} of elements in $S$.  As such, this concept bears a direct relationship with the role of centrality underlying the more traditional mean.

Because there are several types of similarity indices (e.g.~\cite{Steinley,Cha,Wolda,Hamers,CostaSimilarity}), an equal number of similarity means can be obtained, each of these inheriting the properties of the similarity index from which they derive.   

The present work focuses on the means defined from the Jaccard similarity index, presented in Section~\ref{sec:definition}.

Among several alternative similarity index, the Jaccard approach represents a particularly intuitive and powerful manner to compare two non-empty sets.  Basically, this index quantifies how much the two sets intersect one another, normalized by their union.  Therefore, the maximum value of 1 is obtained whenever the two sets are identical, while the minimum value of 0 results when the two sets have no common elements (no intersection).

Interestingly, the Jaccard index can be extended to multisets, and then to virtually any mathematical structure, including vectors, functions, matrices, fields, graphs, etc.~\cite{CostaJaccard,CostaSimilarity,CostaMNeurons}.  Thus, the Jaccard index provides an effective approach to be considered in several scientific and technological areas, including but not limited to signal and image processing (e.g.~\cite{oppenheim1978,gonzalez2009,roberts1987}), pattern recognition (e.g.~\cite{fukunaga1993,duda2000pattern,theodoridis2006pattern}), and artificial intelligence (e.g.~\cite{winston1992,mitchell1997}).

The main motivation to consider the Jaccard index as the basis for defining a similarity mean derives from the important properties that this specific index has been found to present (e.g.~\cite{CostaJaccard,CostaSimilarity,CostaMNeurons}).  These include the
fact that it is intrinsically normalized, taking values in the interval $[0,1]$, being also non-dimensional, highly selective and sensitive, while being robust to moderate perturbations of the data to be compared, as well as to the presence of outliers.

\section{The Jaccard Mean is Contained in the Original Values}

One of the most interesting properties of the Jaccard similarity means corresponds to the fact, as shown in the following, that it necessarily coincides with one of the original values.   This property is especially important because it simplifies the calculation of the means (only the original values need to be checked for maximum similarity), and also because it does not require the support of the random variable to be extended in order to incorporate a new value.

As the indices (order) of the sample values are immaterial, to simplify the argument it is assumed that the $N$ sample values are sorted in non-decreasing order, such that $x_j\le x_{j+1}$ for $1\le j\le N-1$. Now, consider the case $x\le x_1$. The minima in the numerator of Eq.~\eqref{eq:jackx} are always $x$, while the maxima in the denominator are always $x_i$, resulting in
\begin{align}
  \mathcal{J}(x, \pmb{x}) = \frac{N x}{s_N} = \frac{x}{\mu}\le \frac{x_1}{\mu} = \mathcal{J}(x_1, \pmb{x}), \label{eq:discsmall}
\end{align}
where the notation $s_k = \sum_{i=1}^k x_i$ and $\mu = s_N/N$ is used.
On the other hand, for the case $x\ge x_N$ the minima in numerator of Eq.~\eqref{eq:jackx} are always $x_i$ and the maxima in the denominator are always $x$, such that
\begin{align}
  \mathcal{J}(x, \pmb{x}) = \frac{s_N}{N x} = \frac{\mu}{x} \le \frac{\mu}{x_N} = \mathcal{J}(x_N, \pmb{x}). \label{eq:disclarge}
\end{align}
This shows that the maximum of the Jaccard similarity is contained within the interval defined by the sample values, and it remains to be shown that it is reached at one of them.

For $x_1 < x < x_N$, there will be a $k$, $1\le k < N$, for which $x_k \le x < x_{k+1}$. The minima in the numerator are $x_i$ for any $i\le k$ and $x$ for the remaining $N-k$ values; the maxima in the denominator are $x$ for $i\le k$ and $x_i$ for the remaining elements, resulting in
\begin{align}\label{eq:jacksum}
  \mathcal{J}(x,\pmb{x}) = \frac{s_k + (N-k)x}{k x + (s_N - s_k)}.
\end{align}
It is easy to see that Eq.~\eqref{eq:jacksum} is valid also for $x$ assuming both limit values $x_k$ and $x_{k+1}$. Therefore, the derivative
\begin{align}
  \frac{d}{dx}\mathcal{J}(x,\pmb{x}) = \frac{N(s_N-s_k)-k s_N}{\left(k x + (s_N - s_k)\right)^2}
\end{align}
is zero only if
\begin{align*}
  N(s_N-s_k) = k s_N,
\end{align*}
which is independent of $x$. The Jaccard similarity in the interval between two samples is then either growing, with maximum at $x_{k+1}$, decreasing with maximum at $x_k$ or constant. With this it is clear that the Jaccard similarity for values of $x_k < x < x_{k+1}$ cannot be larger than the ones at the sample values $x_k$ and $x_{k+!}$. In the special case where the derivative is zero, there is a degeneracy and it is possible to choose either $x_k$, $x_{k+1}$, or any value in the interval as the mean.

As an example, consider the set of values:
\begin{align}
  S = \left\{ 3, 4, 15, 16 \right\}
\end{align}

As the Jaccard similarity mean of this set, $\tilde{x}$, necessarily belongs to $S$, only its four values need to be considered.  The Jaccard similarities between each of the four candidate elements and the set of values can thus be calculated as follows:
\begin{align*}
  J(3, S) & = \frac{\min(3,3)+\min(3,4)+\min(3,15)+\min(3,16)}{\max(3,3)+\max(3,4)+\max(3,15)+\max(3,16)} = \frac{6}{19} \\
  J(4, S) &= \frac{\min(4,3)+\min(4,4)+\min(4,15)+\min(4,16)}{\max(4,3)+\max(4,4)+\max(4,15)+\max(4,16)} = \frac{15}{39} \\
  J(15, S) &= \frac{\min(15,3)+\min(15,4)+\min(15,15)+\min(15,16)}{\max(15,3)+\max(15,4)+\max(15,15)+\max(15,16)} = \frac{37}{61} \\
  J(16, S) &= \frac{\min(16,3)+\min(16,4)+\min(16,15)+\min(16,16)}{\max(16,3)+\max(16,4)+\max(16,15)+\max(16,16)} = \frac{19}{32}
\end{align*}

Therefore, the Jaccard similarity mean of the set $S$ above
corresponds to the value $\tilde{x}=15$, as this yields the maximum Jaccard similarity value $37/61$ with the set $S$.

Though simple, this example shows that, even if the sough Jaccard mean is known to be contained in the set $S$, a total of $N^2$ minimum and maximum operations will still be required for its calculation, implying $\mathcal{O}(N^2)$.  

The main reason why the Jaccard mean is more computationally expensive to calculate than the arithmetic mean (which is $\mathcal{O}(N)$) is that these similarity indices are intrinsically non-linear, while the traditional average is linear.

In the next section, an algorithm is developed which allows the Jaccard mean to be calculated in $\mathcal{O}(N \log N)$.

\section{Computing the Jaccard Mean}\label{sec:computing}

The above results suggest an efficient algorithm for the computation of the Jaccard mean of a sample of values.  First, notice that the sign of the derivative of the Jaccard similarity is given by the sign of
\begin{align*}
\varsigma(k) = N(s_N - s_k) - k s_N 
\end{align*}
But in this expression, the first term starts as a large positive value of $N(s_N - s_1)$ and decreases monotonically with $k$ to zero, while the second term increases monotonically with $k$ from $s_N$ to $N s_N$. The result is that the derivative starts positive and decreases monotonically to a negative value. It is possible, therefore, to compute the Jaccard mean by the following $\mathcal{O}(N \log N)$ procedure:
\begin{enumerate}
    \item Sort the sample values in non-decreasing order. Call $x_i$ the $i$-th value in this order.
    \item Compute $s_i = \sum_{j=1}^{i}x_j$ for $1\le i\le N$.
    \item Starting at $i=1$, keep computing $\varsigma(i)$ for successive values of $i$ while this expression is positive.
    \item When you find an $m$ for which $\varsigma(m)$ is either negative or zero, use $x_m$ as the Jaccard mean.
\end{enumerate}
In this definition, for the degenerate case where $\varsigma(m)=0$, the smallest value $x_m$ is used, where $x_{m+1}$ could have been used as well. This is done for compatibility with the definition for continuous random variables, to be presented below, as in this case $x_m$ is the sample value with zero derivative.

Besides numerically computing the Jaccard mean by using the above approach, it is also interesting to obtain analytical expressions for several continuous probability distributions, which is presented in the following section. For a brief treatment of discrete distributions, see Appendix~\ref{app:discrete}.

\section{Continuous Distributions}\label{sec:continuous}

As addressed in the present section, continuous probability distributions can also often have their Jaccard mean expressed in analytical terms.

First rewrite Eq.~\eqref{eq:jcontdef} as
\begin{align*}
  \mathcal{J}_X(x) & =  \frac{\int_0^x \min\{x,\xi\}\rho_X(\xi)\,d\xi+\int_x^\infty \min\{x,\xi\}\rho_X(\xi)\,d\xi}{\int_0^x \max\{x,\xi\}\rho_X(\xi)\,d\xi+\int_x^\infty \max\{x,\xi\}\rho_X(\xi)\,d\xi} \nonumber\\
  & = \frac{\int_0^x \xi \rho_X(\xi)\,d\xi+\int_x^\infty x \rho_X(\xi)\,d\xi}{\int_0^x x \rho_X(\xi)\,d\xi+\int_x^\infty \xi \rho_X(\xi)\,d\xi}.\nonumber
\end{align*}
Using the notation
\begin{align}
    P_X(x) & = \int_0^x \rho_X(\xi)\,d\xi,\label{eq:cdf}\\
    m_X(x) & = \int_0^x \xi \rho_X(\xi)\,d\xi,\label{eq:mx}\\
    \mu_X  & = \int_0^\infty \xi \rho_X(\xi)\,d\xi,\label{eq:mean}
\end{align}
results in
\begin{align}
  \mathcal{J}_X(x) = \frac{\left(1 - P_X(x)\right) x + m_X(x)}{P_X(x) x + \mu_X - m_X(x)}.\label{eq:jcont}
\end{align}

Notice that, if the PDF is such that $\rho_X(x)=0$ for $0<x<a$, then
\begin{align}
  \mathcal{J}_X(x) = \frac{x}{\mu_X}, \quad 0\le x\le a,\label{eq:jsmall}
\end{align}
which means that before $x=a$, $J_X(x)$ grows linearly. This is similar to Eq.~\eqref{eq:discsmall}. On the other hand, if the PDF is such that $\rho_X(x)=0$ for $x>b$, then
\begin{align}
  \mathcal{J}_X(x) = \frac{\mu_X}{x},\quad x \ge b,\label{eq:jlarge}
\end{align}
and $J_X(x)$ decays with $1/x$ after $x=b$. This is similar to Eq.~\eqref{eq:disclarge}.

Also, $J_X(a)=a/\mu_X$ and $J_X(b)=\mu_X/b$ are only equal if $\mu_X = \sqrt{a b}$, which is only possible if $a=b$. And because in general $\mu_X>\sqrt{a b}$, $J_X(a) < J_X(b)$, and the Jaccard similarity function if asymmetrical in $a<x<b$, even if $\rho_X(x)$ is symmetrical.

To find the Jaccard mean the derivative of Eq.~\eqref{eq:jcont},
\begin{align*}
    \frac{d}{d x}\mathcal{J}_X(x) = \frac{\left(1-P_X(x)\right)\mu_X-m_X(x)}{\left(P_X(x) x + \mu_X - m_X(x)\right)^2},
\end{align*}
is computed and equated to zero. The value $x$ that solve the equation
\begin{align}
    \varsigma_X(x) =\left(1-P_X(x)\right)\mu_X-m_X(x) = 0 \label{eq:jmean}
\end{align}
is the Jaccard mean of $X$, denoted here $\tilde{x}$. If $\rho_X(x)=0$ in a finite region and it happens that simultaneously $\varsigma_X(x)=0$, then all values of $x$ in this region are solutions to Eq.~\eqref{eq:jmean} and the Jaccard mean is degenerate.

The factor $1-P_X(x)$ decreases (non necessarily monotonically) as $x$ grows, while $m_X(x)$ grows (non necessarily monotonically) with $x$. Also, $\left(1-P_X(x)\right)\mu_X$ is initially $\mu_X$ and drops to 0 at infinity, while $m_X(x)$ is initially zero and grow to $\mu_X$ at infinity. So, if $\rho_X(x)$ is continuous, the derivative will necessarily be zero somewhere. Furthermore, $1-P_X(x)$ and $m_X(x)$ are constant only in regions where $\rho_X(x)=0$. This implies that
\begin{enumerate}
    \item if $\varsigma_X(x) > 0$, then $\tilde{x} > x$, and
    \item if $\varsigma_X(x) < 0$, then $\tilde{x} < x$.
\end{enumerate}

\section{Some Examples}
\label{sec:examples}

Having developed a general analytical expression for the Jaccard mean of continuous distributions, it is now interesting to look at some examples. Here the following distributions are considered: uniform, exponential, truncated normal and power law.

\paragraph{Uniform distribution}

Consider a random variable $U$ with a uniform distribution between $\mu-d$ and $\mu+d$, with $\mu>0$ and $d<\mu$, that is
\begin{align}
    \rho_U(x) = 
    \begin{cases}
         \frac{1}{2d} & \mu-d < x < \mu+d,  \\
         0 & \mbox{otherwise}, 
    \end{cases}
    \label{eq:uniform}
\end{align}
The Jaccard mean of this distribution is given by (see Appendix~\ref{app:continuous}
for details)
\begin{align}
    \tilde{x} = \left(\sqrt{4+\frac{d^2}{\mu^2}}-1\right)\mu.\label{eq:meanuni}
\end{align}
This is always larger than the first moment $\mu$, being equal only when $d/\mu \rightarrow 0$. As, to include only positive values, $d$ has to be at most $\mu$, the largest possible value of $\tilde{x}/\mu$ is $\sqrt{5}-1$. 

\begin{figure}
\centering
\begin{subfigure}{0.4\textwidth}
    \includegraphics[width=\textwidth]{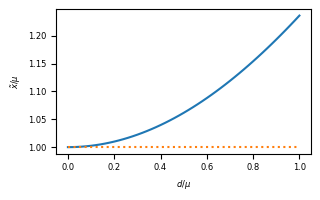}
    \caption{}
    \label{fig:mean-uni}
\end{subfigure}
\begin{subfigure}{0.4\textwidth}
    \includegraphics[width=\textwidth]{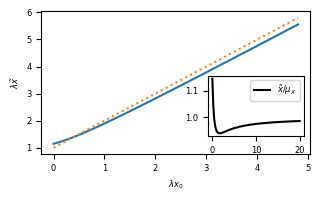}
    \caption{}
    \label{fig:mean-exp}
\end{subfigure}
\begin{subfigure}{0.4\textwidth}
    \includegraphics[width=\textwidth]{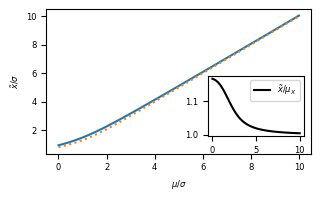}
    \caption{}
    \label{fig:mean-norm}
\end{subfigure}
\begin{subfigure}{0.4\textwidth}
    \includegraphics[width=\textwidth]{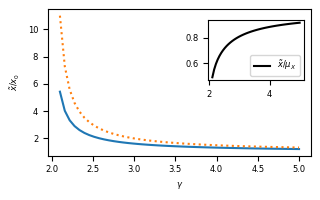}
    \caption{}
    \label{fig:mean-pow}
\end{subfigure}
\caption{The Jaccard mean $\tilde{x}$ for some distributions: (a)~The ratio $\tilde{x}/\mu$ as a function of $d/\mu$ for the uniform distribution of Eq.~\eqref{eq:uniform}; (b)~The value of $\lambda \tilde{x}$ as a function of $\lambda x_0$ for the exponential distribution of Eq.~\eqref{eq:exponential}; the inset shows the ratio of the Jaccard mean to the arithmetic mean as a function of $\lambda x_0$; (c)~The value of $\tilde{x}/\sigma$ as a function of $\mu/\sigma$ for the normal distribution of Eq.~\eqref{eq:trunnorm}; the inset shows the ratio of the Jaccard mean to the arithmetic mean as a function of $\mu/\sigma$; and (d)~The value of $\tilde{x}/x_0$ as a function of $\gamma$ for the power law distribution of Eq.~\eqref{eq:pow}; the inset shows the ratio of the Jaccard mean to the arithmetic mean as a function of $\gamma$. Continuous lines show the Jaccard mean, while the arithmetic mean (dotted line) is included for comparison.}
\label{fig:jmean}
\end{figure}

Figure~\ref{fig:mean-uni} depicts the Jaccard means for the uniform distribution. It is interesting to note the growth of the Jaccard mean with respect to the arithmetic mean as the distribution gets wider.

\paragraph{Exponential Distribution}

Consider an exponential random variable $E$ with PDF 
\begin{align}
    \rho_E(x) = \begin{cases}
               \lambda e^{-\lambda (x-x_0)}, & x>x_0,\\
               0 & x < x_0
            \end{cases} \label{eq:exponential}
\end{align}
The Jaccard mean of such a distribution is (see Appendix~\ref{app:continuous} for details)
\begin{align}
    \tilde{x} = -\left[2+\lambda x_0 + W_{-1}\left(-(1+\lambda x_0)e^{-2(1+\lambda x_0)}\right)\right]\frac{1}{\lambda},\label{eq:meanexp}
\end{align}
where $W_{-1}(x)$ is the -1 branch of the Lambert $W$ function.

Figure~\ref{fig:mean-exp} shows the Jaccard mean for the exponential distribution. As shown, the Jaccard mean is larger than the arithmetic mean for small $\lambda x_0$, but this is soon reverted.

\paragraph{Truncated Normal}

A normal distribution always has support in $x<0$. Instead a random variable $N$ distributed as a truncated normal is used here, with PDF
\begin{align}
    \rho_N(x) = \begin{cases}
        \frac{1}{\alpha} e^{-\frac{(x-\mu)^2}{2\sigma^2}} & x \ge 0,\\
        0 & x < 0,
    \end{cases}\label{eq:trunnorm}
\end{align}
where the normalization constant $\alpha$ is
\begin{align*}
    \alpha = \sigma\sqrt{\frac{\pi}{2}}\left(1+\erf\left(\frac{\mu}{\sqrt{2}\sigma}\right)\right),
\end{align*}
with $\erf(x)$ the error function. For this PDF Jaccard mean is the solution of the equation (See Appendix~\ref{app:continuous}, for details):
\begin{align}
    \sqrt{\frac{2}{\pi}}\frac{\alpha}{\sigma}e^{-\frac{(x-\mu)^2}{2\sigma^2}} & - \left(2\frac{\mu\alpha}{\sigma^2}+e^{-\frac{\mu^2}{2\sigma^2}}\right)\erf\left(\frac{x-\mu}{\sqrt{2}\sigma}\right)\nonumber\\ & - \left(\sqrt{\frac{2}{\pi}}\frac{\alpha}{\sigma}-1\right)e^{-\frac{\mu^2}{2\sigma^2}}+\left(2-\sqrt{\frac{2}{\pi}}\frac{\alpha}{\sigma}\right)\frac{\mu\alpha}{\sigma^2} = 0. \label{eq:meannorm}
\end{align}
This equation can be solved numerically for given values of $\mu$ and $\sigma$. 

Figure~\ref{fig:mean-norm} shows the Jaccard mean for this distribution. It can be seen that, if $\mu$ is sufficiently larger than $\sigma$, such that the truncation of negative values does not significantly affect the distribution, there is a linear relationship from $\tilde{x}/\sigma$ to $\mu/\sigma$, with a ratio close to one for large $\mu/\sigma$.

\paragraph{Power Law}

Consider a random variable $P$ with the following distribution
\begin{align}
    \rho_P(x) = \begin{cases}
        \frac{\gamma-1}{x_0} \left(\frac{x}{x_0}\right)^{-\gamma}& x \ge x_0,\\
        0 & x < x_0,
    \end{cases}\label{eq:pow}
\end{align}
with $x_0>0$ and $\gamma>2$. For this PDF, in the region of interest, $x > x_0$, the equation for the mean is (see Appendix~\ref{app:continuous} for details):
\begin{align}
    \left(\frac{x}{x_0}\right)^{-\gamma+2} + \left(\frac{x}{x_0}\right)^{-\gamma+1} = 1. \label{eq:meanpow}
\end{align}
This equation can be solved numerically for given values of $x_0$ and $\gamma$. 

Figure~\ref{fig:mean-pow} shows the Jaccard mean for this distribution. The Jaccard mean is in this case smaller than the arithmetic mean, specially for smaller values of $\gamma$, which means it is less affected by the large values generated by a power law distribution with small exponent.

\section{Quantifying variability}
\label{sec:variability}

Frequently, in addition to finding a representative value of a dataset, one also wants to quantify the variability of that data around this value. For the arithmetic mean it is customary to use the standard deviation, or its normalized version, the coefficient of variation.

It is possible to readily derive a variation coefficient associated with the Jaccard mean by taking into consideration that the index $\mathcal{J}_X(x)$ quantifies the similarity of the value $x$ with the data as a whole, allied to the fact that the Jaccard mean $\tilde{x}$ is the most similar value to all data.  First, observe that $\mathcal{J}_X(\tilde{x})$ expresses how much the Jaccard average resembles the values in the set $X$.  As a consequence, the larger the quantity $\mathcal{J}_X(\tilde{x})$, the less disperse the values in the dataset will be. Therefore, taking into account also the fact that $0 \leq \mathcal{J}_X(\tilde{x}) \leq 1$,  it is possible to define a \emph{Jaccard variability coefficient} as
\begin{align}
    \upsilon = 1 - \mathcal{J}_X(\tilde{x}). \label{eq:var}
\end{align}
with $0 \leq \upsilon \leq 1$.

When all the values are similar, $\mathcal{J}_X(\tilde{x})$ is close to 1, and therefore $\upsilon$ approaches zero, indicating low data variability; when the values spread widely, the value of $\mathcal{J}(\tilde{x})$ approaches zero, and $\upsilon$ becomes close to one, indicating high variability.  In addition, given that $\upsilon$ is inherently non-dimensional, an analogy can be postulated between that quantity and the traditional coefficient of variation, defined as
\begin{align}
   C_X = \frac{\sigma_X}{\mu_X}
\end{align}
where $\mu_X$ and $\sigma_X$ stand for the arithmetic mean and standard deviation of the values in the set $X$.

\begin{figure}
\centering
\begin{subfigure}{0.4\textwidth}
    \includegraphics[width=\textwidth]{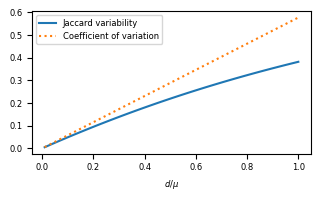}
    \caption{}
    \label{fig:var-uni}
\end{subfigure}
\begin{subfigure}{0.4\textwidth}
    \includegraphics[width=\textwidth]{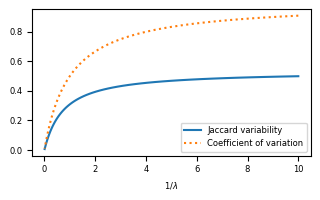}
    \caption{}
    \label{fig:var-exp}
\end{subfigure}
\begin{subfigure}{0.4\textwidth}
    \includegraphics[width=\textwidth]{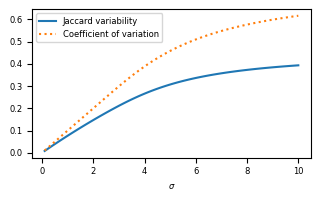}
    \caption{}
    \label{fig:var-norm}
\end{subfigure}
\begin{subfigure}{0.4\textwidth}
    \includegraphics[width=\textwidth]{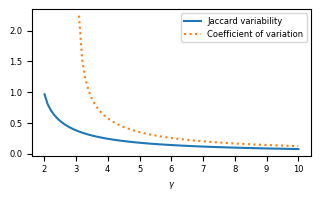}
    \caption{}
    \label{fig:var-pow}
\end{subfigure}
    \caption{The Jaccard variability $\upsilon$ for some distributions: (a)~For the uniform distribution of Eq.~\eqref{eq:uniform}, as a function of $d/\mu$; (b)~For the exponential distribution of Eq.~\eqref{eq:exponential} with $x_0=1$, as a function of $1/\lambda$;  (c)~For the truncated normal distribution of Eq.~\eqref{eq:trunnorm} with $\mu=10$, as a function of $\sigma$; and (d)~For the power law distribution of Eq.~\eqref{eq:pow} with $x_0=1$, as a function of $\gamma$. Continuous lines show the Jaccard variability, while dotted lines show the coefficient of variation, for comparison.}
    \label{fig:var}
 \end{figure}

Figure~\ref{fig:var} shows the Jaccard variability coefficient for our example distributions considered in Section~\ref{sec:examples}, including also the coefficient of variation for comparison. It can be observed that $\upsilon$ grows monotonically with the coefficient of variation, even if non-linearly, which supports its use as a variability metric. The Jaccard variability is normalized between 0 and 1, while the coefficient of variation is unbound. This is clearly seen in the case of the power law distribution of Fig.~\ref{fig:var-pow}, where the coefficient of variation diverges as $\gamma$ decreases to 2.  The fact that the Jaccard variation never diverges constitutes an interesting characteristic of this measurement.

\section{Relationship with the median}
The Jaccard mean $\tilde{x}$ is the solution to Eq.~\eqref{eq:jmean}, and therefore
\begin{align}
    m(\tilde{x}) = \mu\left(1 - P_X(\tilde{x})\right).
\end{align}
Substituting this in Eq.~\eqref{eq:jcont} results in
\begin{align}
    \mathcal{J}(\tilde{x}) = \frac{1-P_X(\tilde{x})}{P_X(\tilde{x})}.
\end{align}
But $\mathcal{J}(x)\le 1$, and therefore
\begin{align}
    P_X(\tilde{x}) \ge \frac{1}{2},
\end{align}
which means that \emph{the Jaccard mean is not smaller than the median}.

\section{Scaling and Translation}

It is know that the arithmetic mean is linearly affected by scaling and translation of the data, such that scaling the data by a factor of $a$ scales the mean by the same factor, and translating the data by a factor $c$ translates the mean by the same factor. It is interesting to evaluate the effect of scaling and translation of the data on the Jaccard mean.  In addition to the intrinsic analytical interest of this analysis, it will also be shown that it is possible to control the bilateral symmetry of the Jaccard average by translating the original data elements.

\paragraph{Scaling}

Given a random variable $X$ with Jaccard mean $\tilde{x}$, a scaled random variable $Y$ created by making $y=ax$ with $a>0$, has
\begin{align*}
    \rho_Y(y) & = \frac{1}{a}\rho_X\left(\frac{y}{a}\right),\\
    P_Y(y) & = P_X\left(\frac{y}{a}\right),\\
    m_Y(y) & = a m_X\left(\frac{y}{a}\right),\\
    \mu_Y  & = a \mu_X,
\end{align*}
and then the Jaccard mean of $Y,$ $\tilde{y},$ is the solution to
\begin{align*}
    \mu_Y\left(1 - P_Y(y)\right) - m_Y(y) & = \\
    a\left[\mu_X\left(1 - P_X\left(\frac{y}{a}\right)\right) - m_X\left(\frac{y}{a}\right)\right] & = 0.
\end{align*}
that is, 
\begin{equation*}
    \tilde{y} = a \tilde{x}.
\end{equation*}

Taking into account Eq.~\eqref{eq:jcont}, it is also easy to see that, in this case
\begin{align}
    \mathcal{J}_Y(y) = \mathcal{J}_X(y/a).
\end{align}

\paragraph{Translation} \label{sec:transation}

Constructing a random variable $Y$ by shifting the values of a given random variable $X$ with Jaccard mean $\tilde{x}$ by $c>0$, that is, $y=x+c$, results, for $y > c$, in
\begin{align*}
    \rho_Y(y) & = \rho_X(y-c)\\
    P_Y(y) & = P_X(y-c)\\
    m_Y(y) & = m_X(y-c) + c P_X(y-c)\\
    \mu_Y  & = \mu_X + c.
\end{align*}
and the Jaccard mean $\tilde{y}$ of $Y$ is the solution to
\begin{align*}
    \mu_Y\left(1 - P_Y(y)\right) - m_Y(y) &= \\
    \mu_X\left(1-P_X(y-c)\right)-m_X(y-c) + c\left(1-2 P_X(y-c)\right) &= 0.
\end{align*}
From this result it can be seen that $\tilde{y} = \tilde{x} + c$ is only possible if $P_X(\tilde{x})=1/2$, that is, if the Jaccard mean of the original distribution is equal to its median $h_X$. Furthermore, as $c$ grows, when it gets much larger than $\mu_X$, the term $c\left(1-2 P_X(y-c)\right)$ dominates the left hand side of the equation, and $\tilde{y} \rightarrow h_X+c = h_Y$ as $c\rightarrow\infty$. Therefore, \emph{by shifting a distribution to the right we bring the Jaccard mean closer to the median}. 

\begin{figure}
\centering
\begin{subfigure}{0.4\textwidth}
    \includegraphics[width=\textwidth]{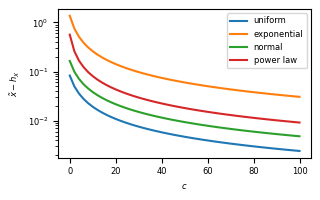}
    \caption{}
    \label{fig:translate-mean}
\end{subfigure}
\begin{subfigure}{0.4\textwidth}
    \includegraphics[width=\textwidth]{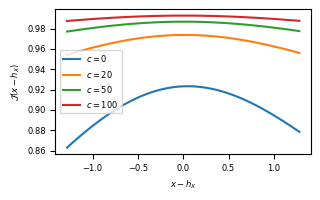}
    \caption{}
    \label{fig:translate-norm}
\end{subfigure}
    \caption{Effects of different translations on the Jaccard mean and similarity: (a)~the difference between the Jaccard mean $\tilde{x}$ and the median $h_X$ of the distribution, for different values of the translation $c$ applied to the following initial distributions: a uniform distribution with $\mu=3$ and $d=1$; an exponential distribution with $\lambda=1/3$ and $x_0=0$; a truncated normal distribution with $\mu=3$ and $\sigma=1$; and a power law distribution with $\gamma=2.5$ and $x_0=1$. All initial distributions have $\mu_X=3$ (in the case of the truncated normal, approximately 3); (b) The similarity in the region from the percentile 10\% to the percentile 90\% for various translations $c$ of an initial truncated normal distribution with $\mu=10$ and $\sigma=1$ (a larger $\mu$ is used here to avoid interference from the truncation); the plot are shifted such that the median $h_X$ is at zero.}
    \label{fig:translate}
 \end{figure}

This can be seen in Fig.~\ref{fig:mean-uni}, using the example of a uniform distribution: For this distribution, a translation corresponds to fixing $d$ and increasing $\mu$, which decreases the value of $d/\mu$, bringing $\tilde{x}/\mu$ closer to one (see Eq.~\eqref{eq:meanuni}). Figure~\ref{fig:translate-mean} shows the effect of a translation in our example distributions of Section~\ref{sec:examples}. In all cases the Jaccard mean gets closer to the median as $c$ grows.

The Jaccard similarity of the translated distribution is
\begin{align}
    \mathcal{J}_Y(x+c) = \frac{\left(1-P_X(x)\right)x + m_X(x) + c}{P_X(x) x + \mu_X - m_X(x) + c},
    \label{eq:jtrans}
\end{align}
that is, a value of $c$ is added in the numerator and denominator of the expression for $J_X(x)$, which increases the value of $\mathcal{J}_Y(y)$ with respect to the corresponding $\mathcal{J}_X(y-c)$. Now, consider a distribution for which most of the density is concentrated in an interval $[a,b]$. In this case, $\mathcal{J}_X(a)\approx a/\mu_X$ and $\mathcal{J}_X(b)\approx\mu_X/b$, but for $Y$ the corresponding values are
\begin{align}
    \mathcal{J}_Y(a+c) &\approx \frac{a + c}{\mu_X + c},\\
    \mathcal{J}_Y(b+c) &\approx \frac{\mu_X + c}{b + c},
\end{align}
which are closer to each other than $\mathcal{J}_X(a)$ and $\mathcal{J}_X(b)$. 

In the special case where the original distribution $p_X(x)$ is symmetric, translation by a large $c$ results in: (i)~getting the maximum of $\mathcal{J}_Y(y)$ (the Jaccard mean) closer to the center of the distribution $h_X+c$, and (ii)~making the values at both sides of the center closer to each other. The result is a symmetrization of the profile of the generally asymmetric Jaccard similarity, at least in the region of high density $[a,b]$. This is shown in Fig.~\ref{fig:translate-norm} for a truncated Gaussian distributions, where the asymmetry of the original Jaccard similarity (corresponding to the $c=0$ curve) is reduced for larger values of $c$. This could be useful in applications where symmetric similarity values are preferred. See more about the effect of translation in Appendix~\ref{app:trans}.

\section{Quantifying the Robustness of the Similarity Mean to Outliers}

Suppose that data is being generated by a mixture of two random processes, a ``reference'' process with PDF $p_R(x)$ and an ``outlier'' process (e.g.~\cite{dixon1953,aggarwal2017,rousseeuw2005}) with PDF $p_O(x)$, and that there is a probability $1-\eta$ of the data coming from the reference process and $\eta$ of coming from the outlier process, resulting in 
\begin{align}
    \rho_X(x) = (1 - \eta)\rho_R(x) + \eta \rho_O(x). \label{eq:pout}
\end{align}

\begin{figure}
    \centering
\begin{subfigure}{0.4\textwidth}
    \includegraphics[width=\textwidth]{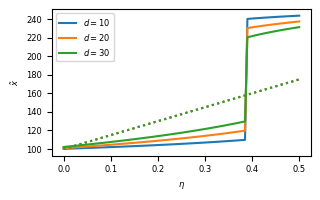}
    \caption{}
    \label{fig:outliers-uni}
\end{subfigure}
\begin{subfigure}{0.4\textwidth}
    \includegraphics[width=\textwidth]{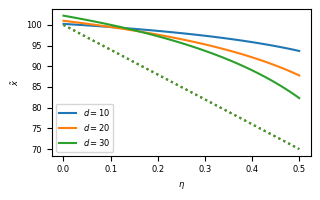}
    \caption{}
    \label{fig:outliers-uni-low}
\end{subfigure}
\begin{subfigure}{0.4\textwidth}
    \includegraphics[width=\textwidth]{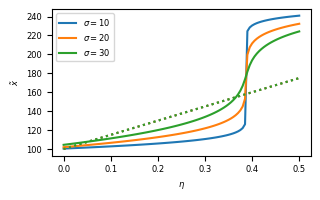}
    \caption{}
    \label{fig:outliers-norm}
\end{subfigure}
\begin{subfigure}{0.4\textwidth}
    \includegraphics[width=\textwidth]{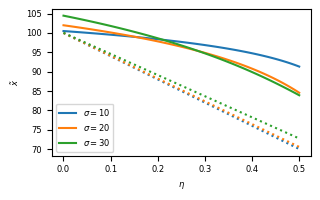}
    \caption{}
    \label{fig:outliers-norm-low}
\end{subfigure}
    \caption{Effect of outliers as a function of the fraction of outliers $\eta$. In (a) and (b) both reference and outlier distributions are uniform, following Eq.~\eqref{eq:uniform}. In (c) and (d) both are truncated normal, following Eq.~\eqref{eq:trunnorm}. The reference distributions always have $\mu=100$. The outlier distributions in (a) and (c) have $\mu=250$, and in (b) and (d) $\mu=40$. The other parameters are shown in the labels of the curves. Solid lines are the Jaccard mean, dotted lines of the same color are the arithmetic means.}
    \label{fig:outliers}
\end{figure}

Figure~\ref{fig:outliers} shows some examples of the effect of outliers on the Jaccard mean. The examples show that, for small values of $\eta$ ($\eta<0.3)$, the Jaccard mean is less affected by outliers than the arithmetic mean: in Figures~\ref{fig:outliers-uni} and~\ref{fig:outliers-norm}, where the outliers are larger than the reference values, the arithmetic mean grows faster with $\eta$ than the Jaccard mean, while in Figures~\ref{fig:outliers-uni-low} and~\ref{fig:outliers-norm-low}, where the outliers are smaller than the reference values the arithmetic mean drops faster with $\eta$ than the Jaccard mean. Larger values of $d$ in Figures~\ref{fig:outliers-uni} and~\ref{fig:outliers-uni-low} or $\sigma$ in Figures~\ref{fig:outliers-norm} and~\ref{fig:outliers-norm-low}, which imply broader distributions of values, increase the effect of outliers on the Jaccard mean. Furthermore, in Figures~\ref{fig:outliers-uni} and~\ref{fig:outliers-norm} a sharp transition of the value of the Jaccard mean to the region of the outliers is seen, occurring for $\eta$ significantly less than $1/2$. This does not happen for $\eta<1/2$ in Figures~\ref{fig:outliers-uni-low} and~\ref{fig:outliers-norm-low}, where the outliers are smaller than the reference values.

To try to shed some light on these results, the following simplifying assumptions are introduced:
\begin{enumerate}
    \item $\eta\approx 0$, that is, outliers are rare;
    \item there is a clear separation between the reference and
      outlier distributions, that is, there is a value $s$ for which:
    \begin{itemize}
        \item for outliers larger than the reference values, $p_R(x)\approx 0$ for $x > s$ and $p_O(x)\approx 0$ for $x < s$; or
        \item for outliers smaller than the reference values, $p_R(x)\approx 0$ for $x < s$ and $p_O(x)\approx 0$ for $x > s$.
    \end{itemize}
\end{enumerate}
These assumptions are used to evaluate the effect of the outliers on the Jaccard mean, that is, on the root of Eq.~\eqref{eq:jmean}.

\paragraph{Large outliers} 
If the outliers are larger than the reference values
\begin{align*}
    P_X(x) & \approx (1-\eta) P_R(x), \\
    m_X(x) & \approx (1-\eta) m_R(x), \\
    \mu_X  & = (1-\eta)\mu_R + \eta \mu_O = \mu_R + \eta (\mu_O - \mu_R).
\end{align*}
Substitution in Eq.~\eqref{eq:jmean} we get
\begin{align*}
    \varsigma_X(x) = (\mu_R + \eta(\mu_O - \mu_R))\left[1 - (1-\eta)P_R(x)\right]-(1-\eta)m_R(x),
\end{align*}
and therefore, under these assumptions, the outlier distribution only affects the Jaccard mean through the outlier probability $\eta$ and the outlier mean $\mu_O$. Rearranging and keeping only linear terms in $\eta$ gives
\begin{align*}
    \varsigma_X(x) = (1-\eta)\varsigma_R(x) +
    \eta\left[\mu_O(1-P_R(x)) + \mu_R P_R(x)\right],
\end{align*}
where $\varsigma_R(x)=\mu_R(1-P_R(x))-m_R(x)$. This expression implies that 
\[\varsigma_X(\tilde{x}_R)=\eta\left[\mu_O(1-P_R(\tilde{x}_R)) + \mu_R P_R(\tilde{x}_R)\right]\]
and at this point it has an slope of
\[\varsigma_X'(\tilde{x}_R)=(1-\eta)\varsigma_R'(\tilde{x}_R)=-(1-\eta)(\tilde{x}_R + \mu_R)p_R(\tilde{x}_R),\] 
from which the approximation (again, keeping only first order in $\eta$)
\begin{align}
    \tilde{x} = \tilde{x}_R + \eta \frac{\mu_O(1-P_R(\tilde{x}_R)) + \mu_R P_R(\tilde{x}_R)}{(\tilde{x}_R + \mu_R)p_R(\tilde{x}_R)}
    \label{eq:outlier-pos}
\end{align}
results.

\paragraph{Small outliers} 
For the case of outliers smaller than the reference values, 
\begin{align*}
    P_X(x) & \approx \eta + (1-\eta) P_R(x), \\
    m_X(x) & \approx \eta\mu_O + (1-\eta) m_R(x), \\
    \mu_X  & = (1-\eta)\mu_R + \eta \mu_O = \mu_R - \eta (\mu_R - \mu_O).
\end{align*}
Substitution in Eq.~\eqref{eq:jmean} gives
\begin{align*}
    \varsigma_X(x) = (1-\eta)(\mu_R - \eta(\mu_R - \mu_O))\left[1 - P_R(x)\right] - \eta\mu_O - (1-\eta)m_R(x).
\end{align*}
Rearranging and keeping only linear terms in $\eta$ results in
\begin{align*}
    \varsigma_X(x) = (1-\eta)\varsigma_R(x) -
    \eta\left[\mu_O P_R(x) + \mu_R (1-P_R(x))\right],
\end{align*}
The slope of $\varsigma_X$ at $\tilde{x}_R$ is
\[\varsigma_X'(\tilde{x}_R)=-(1-\eta)(\tilde{x}_R + \mu_R)p_R(\tilde{x}_R),\] 
but now
\[\varsigma_X(\tilde{x}_R)=-\eta\left[\mu_O P_R(\tilde{x}_R) + \mu_R (1-P_R(\tilde{x}_R))\right]\]
from which the approximation
\begin{align}
    \tilde{x} = \tilde{x}_R - \eta \frac{\mu_O P_R(\tilde{x}_R) + \mu_R (1- P_R(\tilde{x}_R))}{(\tilde{x}_R + \mu_R)p_R(\tilde{x}_R)}
    \label{eq:outlier-neg}
\end{align}
is found.

Expressions \eqref{eq:outlier-pos} and \eqref{eq:outlier-neg} show that the response to outliers is inversely proportional to the density at the Jaccard mean of the undisturbed distribution.  This explains why broader distributions are more sensitive to outliers, as seen in Figure~\ref{fig:outliers}, because broader distributions reduce the values of the density at the mean.

It is easy, given a reference distribution and the average and the fraction of outliers, to use these expressions to determine if the Jaccard mean is more or less sensitive to the outliers than the arithmetic mean.

With regard to the sharp jump in the value of $\tilde{x}$ with $\eta$ seen in Figure~\ref{fig:outliers}, it is not possible to use these expressions, as the jump occurs for a relatively large value of $\eta$. But the reason for the jump is easily seen, as the value of $\varsigma_X(x)$ depends on $P_X(x)$ and $m_X(x)$, and both of these vary very slowly if $\rho_X(x)\approx 0$. Therefore, in the separation region between $\rho_R(x)$ and $\rho_O(x)$, where both have small (or zero) densities, a large variation in $x$ is needed to effect a significant variation in $\varsigma_X(x)$. Thus, as soon as the value of the Jaccard mean is pulled by the outliers outside the high $\rho_R(x)$ region, the mean will be pulled fast all the way up to the region of large $\rho_O(x)$. The transition is faster, the smaller the value of $\rho_X(x)$ in the separation region, and a discontinuity (as seen in Figure.~\ref{fig:outliers-uni}) appears if the density is zero in this region.

\subsection{Outliers and Translation} \label{sec:out-trans}

Looking at Figure~\ref{fig:outliers} and considering the discussion above it is clear that, even if the Jaccard mean is in general less sensitive to outliers than the arithmetic mean,  it can be pulled to the outlier region before $\eta=0.5$ if the outliers are larger and sufficiently frequent. This is not ideal in case there could be a high density of outliers in our sample. But, as seen in Section~\ref{sec:transation}, a positive translation of the values pulls the Jaccard mean in the direction of the median. This can be used to prevent against a too early pull of the Jaccard mean to the outlier region: Before the analysis, the values are translated by a sufficiently large amount, which will pull the Jaccard mean in the direction of the median, which is, for $\eta<0.5$ necessarily in the region of the reference values. 

\begin{figure}
    \centering
\begin{subfigure}{0.4\textwidth}
    \includegraphics[width=\textwidth]{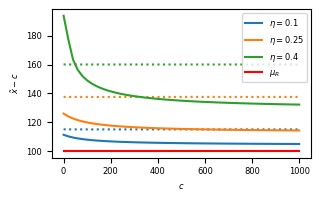}
    \caption{}
    \label{fig:outliers-trans-high}
\end{subfigure}
\begin{subfigure}{0.4\textwidth}
    \includegraphics[width=\textwidth]{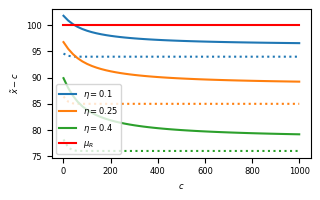}
    \caption{}
    \label{fig:outliers-trans-low}
\end{subfigure}
    \caption{Effect of translation in the sensitivity to outliers of the Jaccard mean. A mixture distribution is used where reference and outlier distributions are truncated normal. The reference distribution has $\mu=100$ and the outlier distribution has $\mu=250$ in (a) and $\mu=40$ in (b). All distributions have $\sigma=30$. The original distributions are translated by $c$, the Jaccard and arithmetic means are calculated and $c$ is subtract from the results to facilitate the comparison. The red lines show the value of the  arithmetic mean of the reference distribution.}
    \label{fig:outliers-trans}
\end{figure}

This is illustrated in Figure~\ref{fig:outliers-trans}. In this figure, the same reference and outlier distributions as in Figure~\ref{fig:outliers-norm} and Figure~\ref{fig:outliers-norm-low} are used, but before computing the mean a translation of the distributions by a variable amount $c$ is done. Afterwards, to facilitate the comparison, $c$ is subtracted from the computed mean. The case of most interest is depicted in Figure~\ref{fig:outliers-trans-high}, where the outliers have large values. It is seen that for this case a value of $c$ around five times the arithmetic mean is able to reduce the value of the (translated) Jaccard mean, pulling it back to the region of the reference distribution, even for high $\eta$. Although Figure~\ref{fig:outliers-trans-low} shows that the translation does not help when the outliers have small values, it also shows that even in this case the (translated) Jaccard mean is less affected by outliers than the arithmetic mean. This shows that translation can be safely used even if there is a possibility of small value outliers.

\section{Identifying Skewness in Datasets}  \label{sec:method}

The fact that the similarity means presents greater robustness to outliers paves the way to a methodology for identification of possible bias in a data set, and for obtaining a more stable quantification of centrality previously suggested in~\cite{CostaSimMean}.

The basic principle is that a dataset in presence of skewed perturbations will have its arithmetic mean $\bar{x}$ to be noticeably distinct from the Jaccard mean $\tilde{x}$.  The congruence between these two means can be quantified, for instance, in terms of the following ratio:
\begin{align}
  \kappa = \frac{\bar{x} - \tilde{x}}{|\tilde{x}| + 1}  \label{eq:kappa}
\end{align}
where the `+1' in the denominator accounts for the eventual case $\tilde{x}=0$.

The methodology then consists of, given a dataset, calculating its arithmetic and similarity means and obtaining the coefficient $\kappa$ by using the above equation.  The case in which $\kappa$ is large can then be taken as an indication of possible skewed perturbations (e.g.\ outliers) in the original dataset.  In addition, in these cases the similarity mean provides a more stable indication of the data centrality.

\begin{figure}
    \centering
\begin{subfigure}{0.4\textwidth}
    \includegraphics[width=\textwidth]{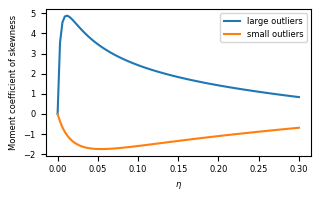}
    \caption{}
    \label{fig:kappaskew}
\end{subfigure}\\
\begin{subfigure}{0.4\textwidth}
    \includegraphics[width=\textwidth]{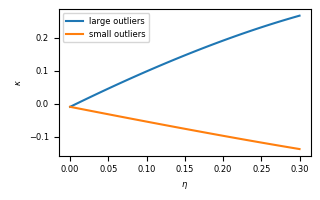}
    \caption{}
    \label{fig:kappakappa}
\end{subfigure}
\begin{subfigure}{0.4\textwidth}
    \includegraphics[width=\textwidth]{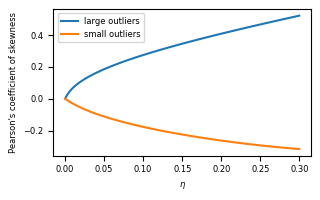}
    \caption{}
    \label{fig:pearsonskew}
\end{subfigure}
    \caption{Effect of the presence of outliers on (a)~the moment coefficient of skewness (i.e.\ the ration between the third central moment and the $3/2$-power of the second central moment) (b)~the $\kappa$ index of Eq.~\eqref{eq:kappa}, and (c)~Pearson's coefficient of skewness (difference between the arithmetic mean and the median, normalized by the standard deviation). The mixture model of Eq.~\eqref{eq:pout} is used with reference and outlier distributions uniform, the reference distribution being uniform in $[80,120]$ and the outlier distribution in $[230, 270]$ for large outlier values or $[20,60]$ for small outlier values.}
    \label{fig:kappa2}
\end{figure}

To evaluate the use of the above defined $\kappa$ as an indicator of the presence of outliers, the mixture model of Eq.~\eqref{eq:pout} is used in Figure~\ref{fig:kappa2}, where it is compared to the moment coefficient of skewness and Pearson's coefficient of skewness. The moment coefficient of skewness is very sensitive to a small amount of outliers (small $\eta$), but when $\eta$ starts to grow (at around 5\% in the presented case), it looses sensitivity. The $\kappa$ index and Pearson's coefficient, on the other hand, have similar behavior, changing monotonically with $\eta$. Therefore, $\kappa$ and Pearson's coefficient are more appropriate than the moment coefficient when a large number of outliers is expected, the choice between the two being depending on the context of the research.

An application of this approach to the study of real-world sunspots is described in the following section.

\section{Application Example:  Sunspots}

The study of the number of sunspots along time has attracted continued interest from the scientific community (e.g.~\cite{bray1964,solanki2011,hathaway1994,clette2014}), not only for its intrinsic interest, but also for its possible effects on Earth.  In particular, sunspots provide indications about possible subsequent solar flares, which can imply severe interference on radio communication and temporarily expand our atmosphere. 

\begin{figure}
\begin{center}
    \includegraphics[width=0.6\linewidth]{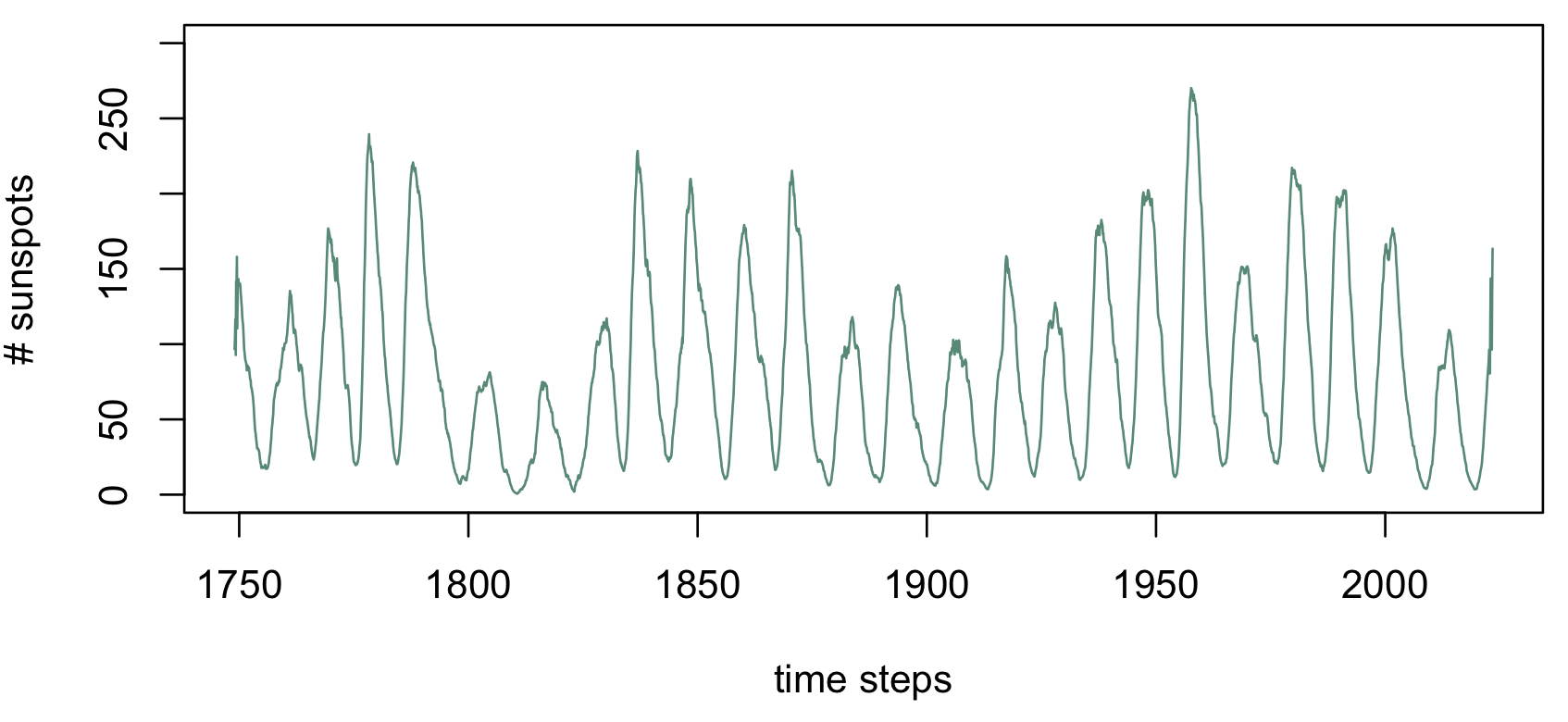}   \\ 
    \caption{Smoothed (moving average with window with length of 19 months) representation of the monthly sequence of the number of sunspots as observed from 1750 to the present (Source: WDC-SILSO, Royal Observatory of Belgium, Brussels~\cite{sidc}).}
    \label{fig:sunspots}
    \end{center}
\end{figure}

Thanks to long term continuing observations, the number of sunspots along time has been found to be nearly periodic, with a period of approximately 11 years defining the \emph{solar cycle} (also known as Schwabe cycle).
Figure~\ref{fig:sunspots} illustrates the sequence of sunspots observed in a monthly basis from 1750 to the present date~\cite{sidc}, smoothed by using moving average with a window with total length of 19 months.

\begin{figure}
\begin{center}
    \includegraphics[width=0.6\linewidth]{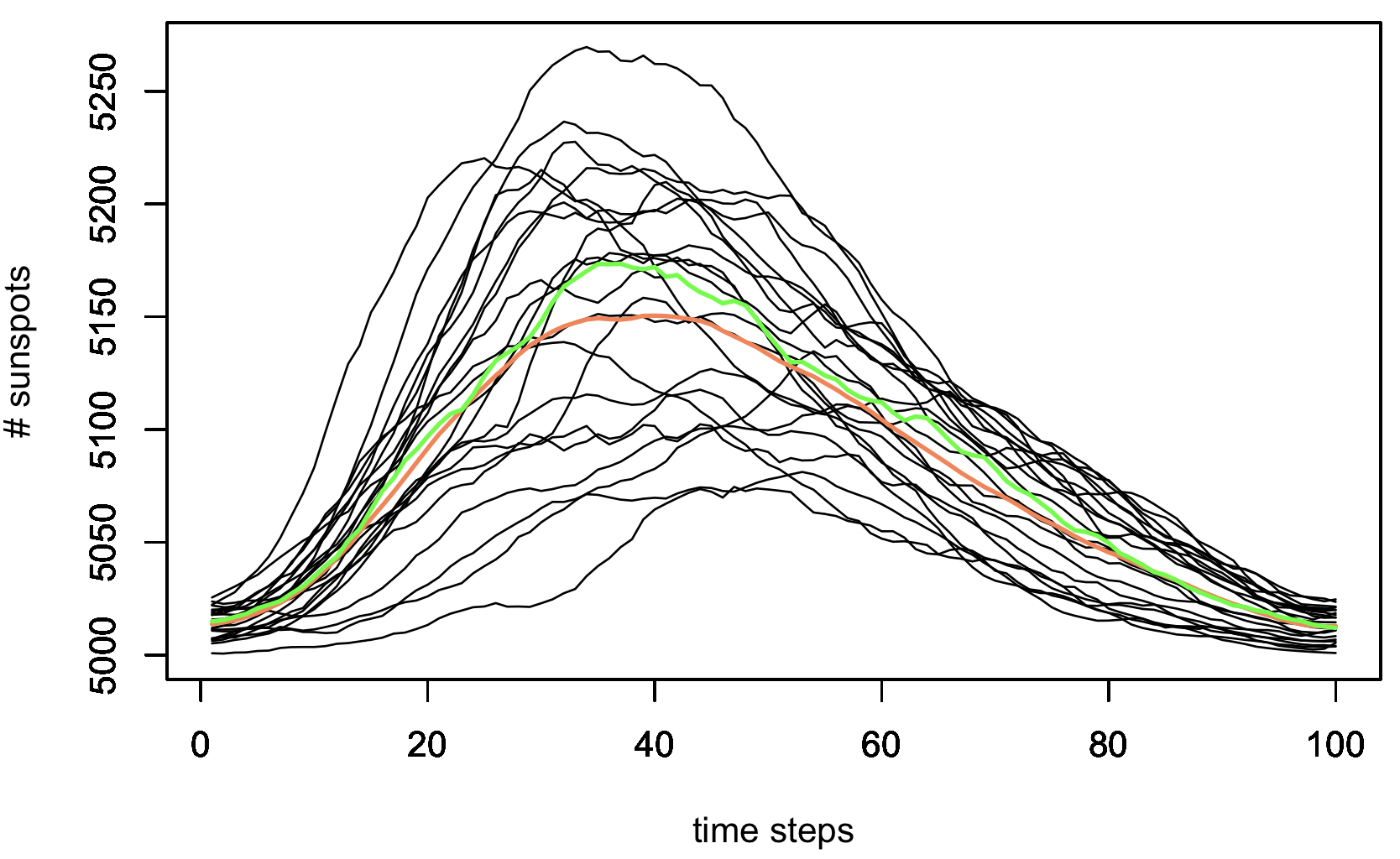}   \\ 
    \caption{Superimposition of the number of sunspot along each solar cycle from 1755 to 2008. The duration of each original solar cycle was linearly interpolated into 100 time steps.  The arithmetic and Jaccard means are shown in orange and green, respectively.   Observe that the number of sunspots in each solar cycle has been added to a constant offset value of 5000 as a means to enhance the symmetry of the Jaccard similarity (see Section~\ref{sec:transation})}
    \label{fig:sun_means}
    \end{center}
\end{figure}

It can be readily observed from Figure~\ref{fig:sunspots} that the numbers of sunspots change considerably along subsequent cycles.  Figure~\ref{fig:sun_means} presents the superimposition of the 24 solar cycles initiating at 1755 up to 2008, which considered the cycles delimitation~\cite{wikisolarcycle} followed by a linear interpolation into 100 time steps.

The sunspots along each of the solar cycles  shown in Figure~\ref{fig:sun_means} indicate that each cycle initiates with a minimum intensity, progresses to a maximum, then returning to minimum activity.  The heights of the peaks can be found to present substantial dispersion that may be a consequence of data skewness.  

In order to obtain further indication about the skewness of the solar cycles in Figure~\ref{fig:sun_means}, we apply the methodology described in Section~\ref{sec:method} considering the arithmetic and Jaccard means taken at each of the 100 interpolated time steps, which are shown as the orange and green curves, respectively.

\begin{figure}
\begin{center}
    \includegraphics[width=0.6\linewidth]{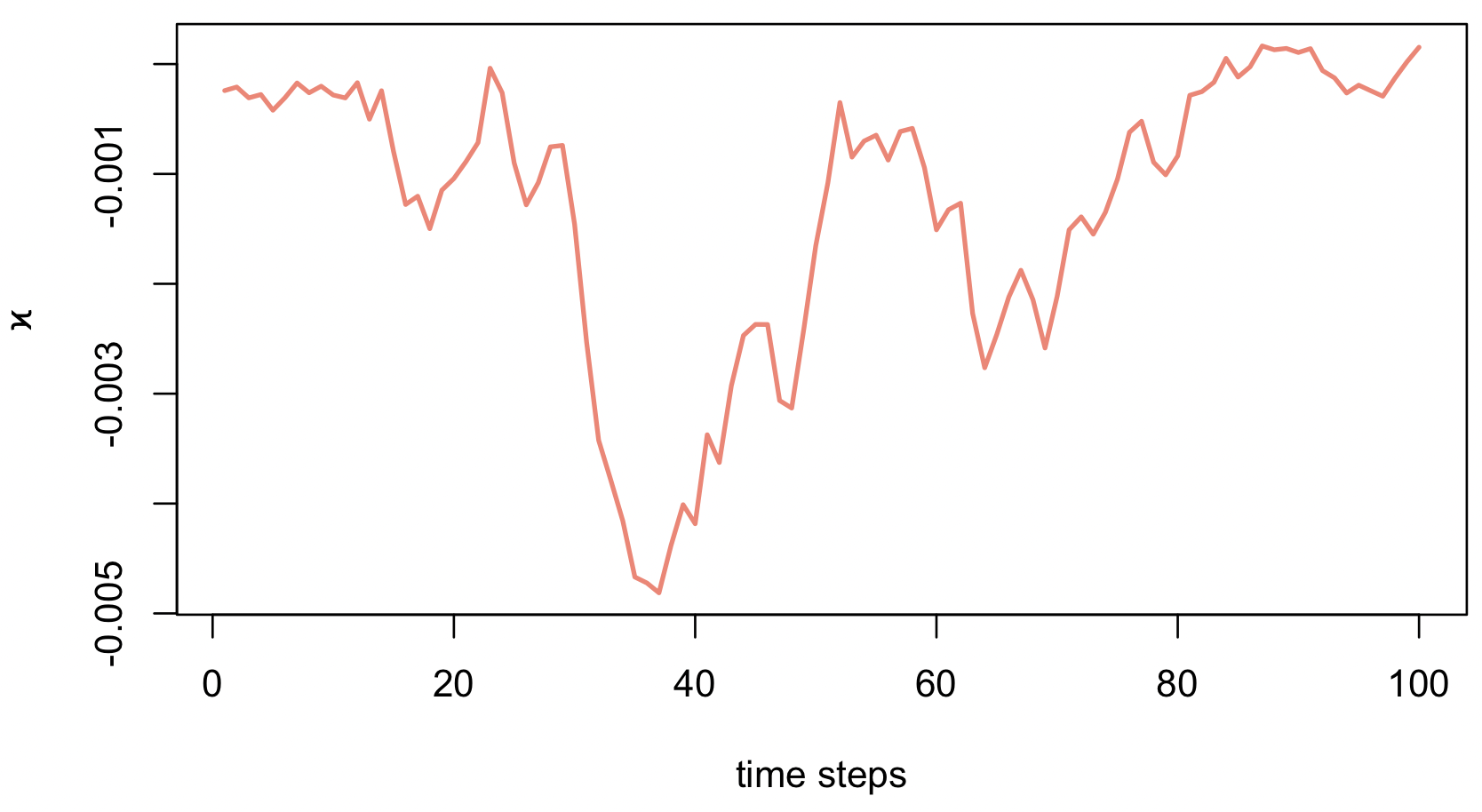}   \\ 
    \caption{The $\kappa$ index between the arithmetic and Jaccard means obtained for the sunspots dataset.  Most of the values are negative, indicating that the arithmetic mean tended to be smaller than the Jaccard counterpart.  A peak of magnitude of $\kappa$ can be observed near time step 40, corresponding to the region most affected by the data skewness.}
    \label{fig:kappa}
    \end{center}
\end{figure}

The fact that the curve corresponding to the arithmetic mean resulted lower than the curve obtained for the Jaccard mean suggests a moderate skewness of the solar cycles data.   This relative displacement between the arithmetic and Jaccard means can be more objectively quantified in terms of the $\kappa$ index in Eq.~\eqref{eq:kappa}.   Figure~\ref{fig:kappa} illustrates the values of this index in terms of the 100 time steps.

The obtained curves of arithmetic and Jaccard means, jointly with the respective $\kappa$ index, suggest the presence of an outlier group of solar cycles with smaller overall numbers of sunspots.  Indeed, a closer inspection of the number of sunspots along each solar cycle confirms that the lower intensity cycles can be indeed understood as an outlier (or skewed) subgroup that ended up generating a multimodal distribution, thus biasing the arithmetic mean away from the Jaccard mean.  No similar subgroups can be found with particularly large numbers of sunspots, which would have otherwise implied an opposite skewness.

\begin{figure}
\begin{center}
    \includegraphics[width=0.6\linewidth]{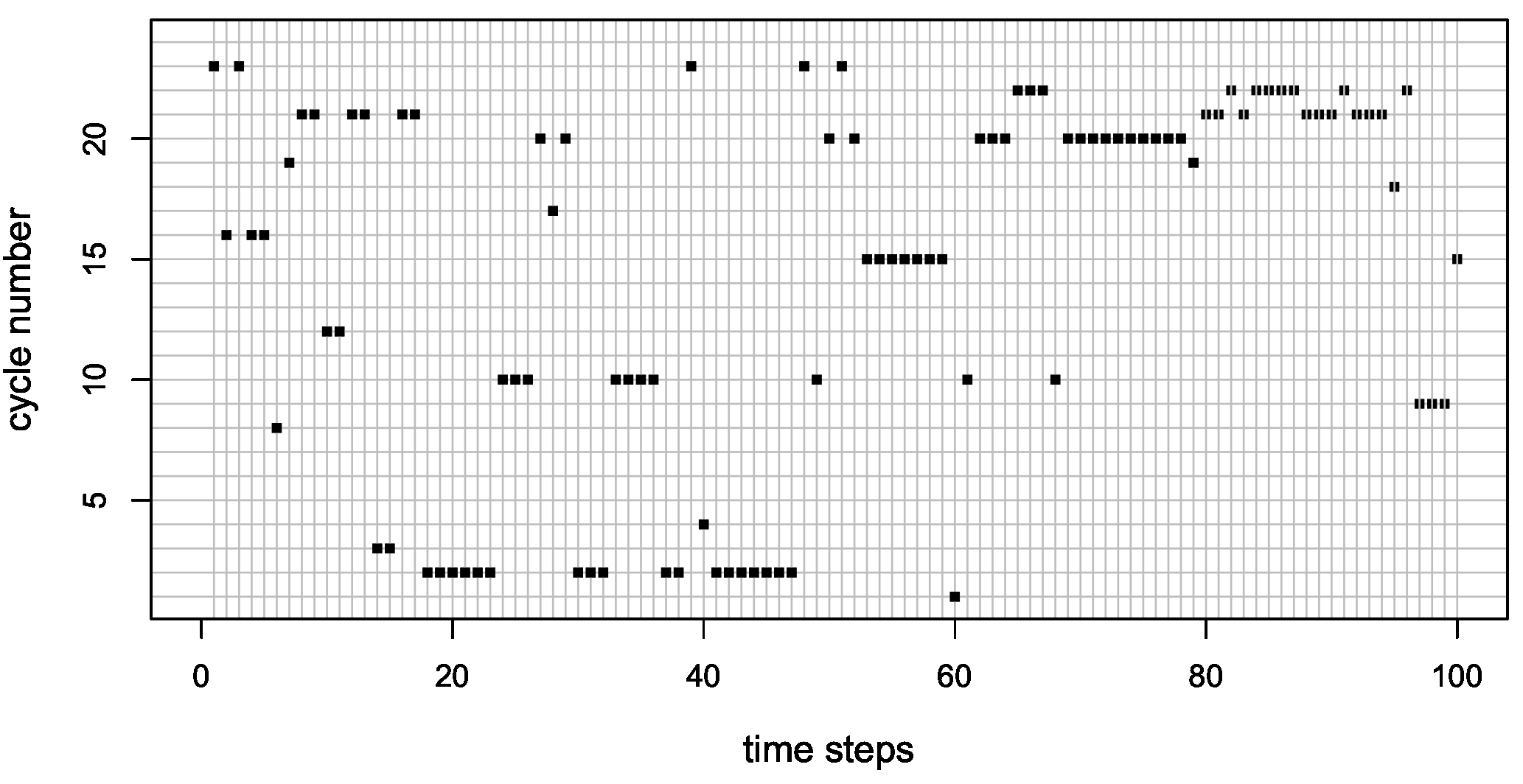}   \\ 
    \caption{Identification of the solar cycle corresponding to the Jaccard mean obtained for each of the 100 time steps.  }
    \label{fig:table_prots}
    \end{center}
\end{figure}

Because the Jaccard mean of a dataset has been shown in this work to necessarily correspond to one of the original data elements, it becomes possible to obtain a table indicating the solar cycle providing the number of sunspots taken as Jaccard means along each of the 100 time steps.  This table is shown in Figure~\ref{fig:table_prots}.

Interestingly, the number of sunspots in the solar cycles 2 (initiating on May 1766) and 10 (Dec.~1855) were chosen as Jaccard means for most of the the time steps between 18 and 47.  Similarly, the number of sunspots from  solar cycles 20 (Sep.~1964), 21 (Feb.~1976), and 22 (Aug.~1986) were taken as Jaccard mean for most of the time steps between 62 and 94.

In other words, the above results indicate that the solar cycles 2 and 10 can be understood as a prototype (or models) of the sunspots along the intermediate time steps, encompassing the region of the solar cycles where the skewness has been identified to be most intense.  In an analogous manner, the solar cycles 20, 21, and 22 can be taken as prototypes of the trailing portion of the respective curves of numbers of sunspots.

\begin{figure}
\begin{center}
    \includegraphics[width=0.6\linewidth]{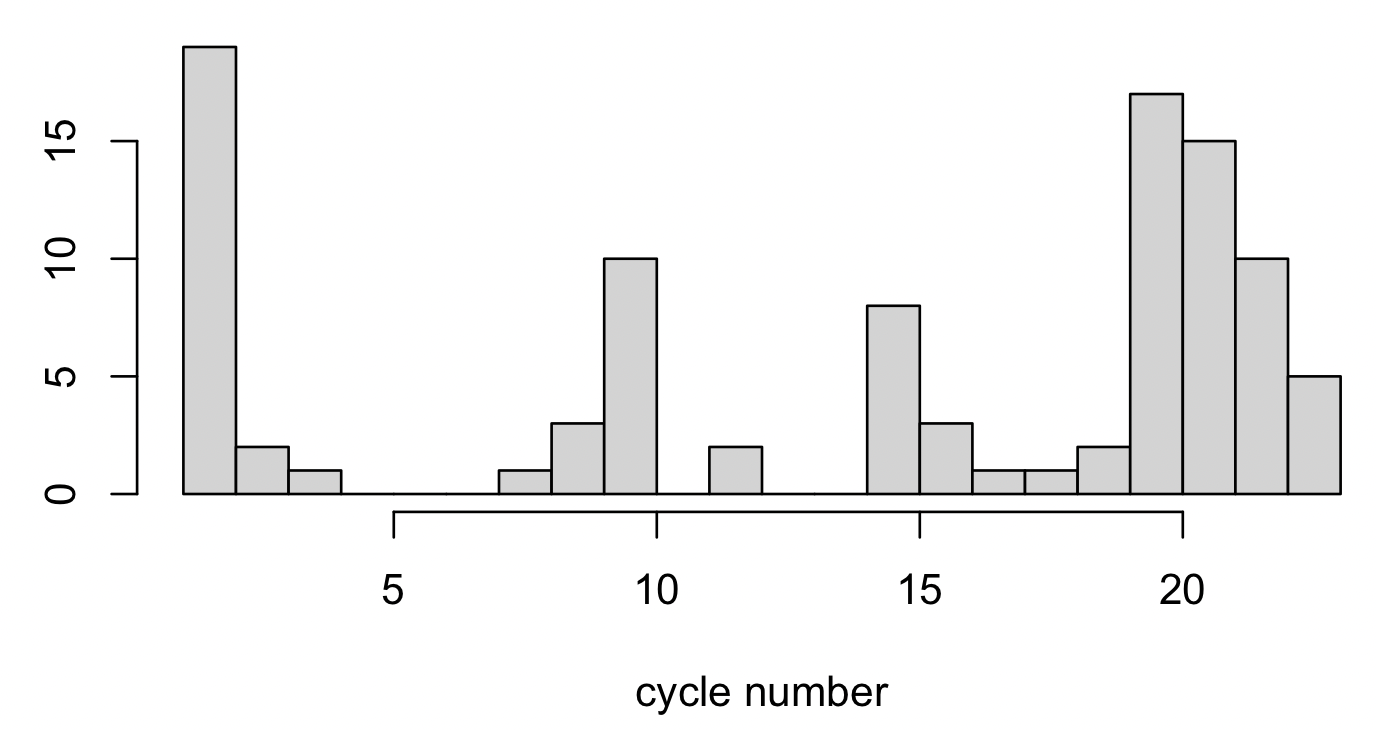}   \\ 
    \caption{Histogram of occurrences of each solar cycle taken as Jaccard mean.  Cycles 2, 10, 20, 21, and 22 are those appearing most frequently along the Jaccard means of the sunspot curves.}
    \label{fig:hist}
    \end{center}
\end{figure}

Figure~\ref{fig:hist} depicts the histogram of occurrences of each solar cycle number taken as Jaccard means along the 100 time steps.  The largest histogram counts are verified for solar cycles 2, 20, 21, 22 and 10.

\section{Concluding Remarks}

The concept of mean plays a critical role not only in the physical sciences, but in almost every quantitative aspect of human activity.  Its importance and ubiquity are implied by its ability to summarize a set of values in terms of single value, indicating the centrality of the values distribution.

While mean has been traditionally associated to arithmetic operations involving the original values to be summarized, the existence of several interesting similarity operators including the Jaccard and coincidence (e.g.~\cite{CostaJaccard,CostaSimilarity})) indices, motivates the definition and characterization of an associated \emph{similarity mean}, which constituted the main objective of the present work.

After reviewing the basic concepts to be used, the concept of similarity means is described and discussed, with emphasis on the Jaccard similarity index, from which the Jaccard mean studied in the present work is derived.   

Several studies and contributions are presented, including the verification that the Jaccard mean of a discrete set of values necessarily corresponds to one of those values, the presentation of an effective algorithm for calculating the Jaccard mean, the analytic determination of the Jaccard mean of several distributions (uniform, exponential, truncated normal, and power law), as well as a study of how the Jaccard mean is influenced by scaling and translation of the original data.   The important issue of robustness to outliers was addressed next, including how it is influenced by translations of the original values.

The potential of the Jaccard mean for data analysis was then illustrated with the study of sunspots cycles.  More specifically, solar cycles were normalized along their domain, and their arithmetic and Jaccard means were estimated for each point along their domain.  The fact that these two means resulted different allowed the identification of a group of outlier solar cycles, characterized by smaller amplitudes.

The development in this text of the concept of the Jaccard mean and the associated variability coefficient provide an additional, complementary tool for data analysis, which is however not meant as a substitutes for more traditional ways to characterize the mean and variability, such as the arithmetic mean and the coefficient of variation. These new resources could be particularly useful when similarity measures are already being employed in the analysis, since they are intrinsically related and compatible with the concept of similarity.

The several reported contributions on concepts and methods related to the Jaccard mean pave the way to a wide range of possible subsequent related investigations.   For instance, the reported results could be extended to data with negative values, and it would be also interesting to characterize and compare similarity means derived from other indices including S{\o}rensen-Dice(e.g.~\cite{carass2020}), overlap (e.g.~\cite{Kavitha}) and coincidence.  Another particularly interesting development would be to extend the obtained results to higher dimensional spaces.

\section*{Acknowledgments}
Luciano da F. Costa thanks CNPq (grant no.~313505/2023-3) and FAPESP
(grants 15/22308-2 and 2022/15304-4).

\bibliographystyle{unsrt}
\bibliography{refs}

\begin{appendices}

\section{Discrete Distributions}\label{app:discrete}

The main text focused on sampled data and on continuous distribution. In this appendix, basic analytic results are derived concerning the Jaccard mean of generic discrete distributions.

Consider a distribution on a set of discrete values $x_i$, $i=1\ldots$, each with probability $p_X(x_i)$, and  assume that $x_i < x_{i+1}$ and that $x_k \le x < x_{k+1}$ for some $k$. In Eq.~\eqref{eq:jaccdist}, 
\begin{align}
  \mathcal{J}_X(x) = \frac{\sum_{i=1}^k x_i p_X(x_i) + x \sum_{i=k+1}^\infty p_X(x_i)}  
                          {x \sum_{i=1}^k p_X(x_i) + \sum_{i=k+1}^\infty x_i p_X(x_i)}.
                          \label{eq:jaccdistint}
\end{align}
Using 
\begin{align*}
    P_X(k) & = \sum_{i=1}^k p_X(x_i),\\
    m_X(k) & = \sum_{i=1}^k x_i p_X(x_i),\\
    \mu_X  & = \sum_{i=1}^\infty x_i p_X(x_i),
\end{align*}
Eq.~\eqref{eq:jaccdistint} becomes
\begin{align}
  \mathcal{J}_X(x) = \frac{m_X(k) + x \left(1 - P_X(k)\right) }  
                          {x P_X(k) + \mu_X - m_X(k) }.
                          \label{eq:jaccdistsimp}
\end{align}
The results of Section~\ref{sec:computing} can be generalized for the computation of the Jaccard mean, notably that the mean is one of the distribution values and that it is the smallest value $x_k$ for which 
\begin{align}\label{eq:signfidv}
\varsigma_X(k) = \mu_X\left(1 - P_X(k)\right) - m_x(k) 
\end{align}
is non-positive.

\section{Computing the Jaccard mean of continuous distributions}\label{app:continuous}

This appendix gives more detail on the computation of the Jaccard mean for the continuous distributions of Section~\ref{sec:examples}.

\subsection{Uniform distribution}

For the random variable $U$ defined by the distribution of Eq.~\eqref{eq:uniform}, focusing in the region of interest between $\mu-d$ and $\mu+d$, Eqs.~\eqref{eq:cdf}, \eqref{eq:mx}, and~\eqref{eq:mean} become
\begin{align*}
    P_U(x) & = \frac{x - \mu + d}{2d},\\
    m_U(x) & = \frac{x^2-(\mu-d)^2}{4d},\\
    \mu_U  & = \mu.\\
\end{align*}
Substituting in Eq.~\eqref{eq:jcont} gives
\begin{align}
    \mathcal{J}_U(x) = \frac{4 d \mu - (x-\mu-d)^2}{4 d \mu + (x-\mu+d)^2}.\label{eq:juni}
\end{align}
Substituting in Eq.~\eqref{eq:jmean} results in
\begin{align*}
    \mu\left(1-\frac{x-\mu+d}{2d}\right)-\frac{x^2-(\mu-d)^2}{4d}=0,
\end{align*}
whose positive solution is Eq~\eqref{eq:meanuni} of the main text.

\begin{figure}
\centering
\begin{subfigure}{0.4\textwidth}
    \includegraphics[width=\textwidth]{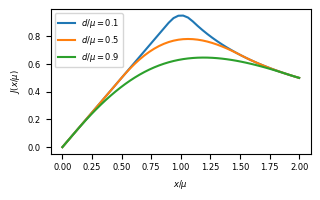}
    \caption{}
    \label{fig:sim-uni}
\end{subfigure}
\begin{subfigure}{0.4\textwidth}
    \includegraphics[width=\textwidth]{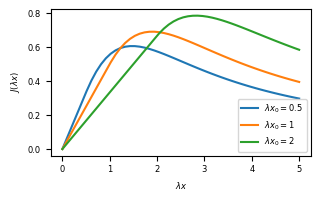}
    \caption{}
    \label{fig:sim-exp}
\end{subfigure}
\begin{subfigure}{0.4\textwidth}
    \includegraphics[width=\textwidth]{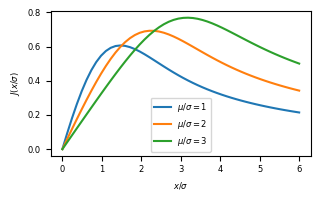}
    \caption{}
    \label{fig:sim-norm}
\end{subfigure}
\begin{subfigure}{0.4\textwidth}
    \includegraphics[width=\textwidth]{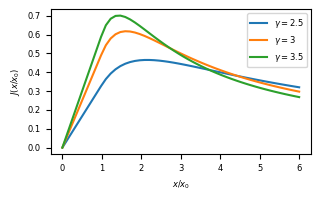}
    \caption{}
    \label{fig:sim-pow}
\end{subfigure}
    \caption{The Jaccard similarity for: (a)~the uniform distribution of Eq.~\eqref{eq:uniform}, as a function of $x/\mu$, for some values of $d/\mu$; (b)~for the exponential distribution of Eq.~\eqref{eq:exponential}, as a function of $\lambda x$ for some values of $\lambda x_0$; (c)~the truncated normal distribution of Eq.~\eqref{eq:trunnorm}, as a function of $x/\sigma$, for some values of $\mu/\sigma$; (d)~the power law distribution of Eq.~\eqref{eq:pow}, as a function of $x/x_0$, for some values of $\gamma$.}
    \label{fig:sim-examples}
 \end{figure}

Figure~\ref{fig:sim-uni} depicts the Jaccard similarity for the uniform distribution. It is noticeable the lack of asymmetry of the similarity, even for this symmetric distribution. Also seen is a linear growth of the similarity before $\mu-d$ and a $1/x$ decay after $\mu+d$, as expected by Eq.~\eqref{eq:jsmall} and \eqref{eq:jlarge}.

\subsection{Exponential Distribution}

The exponential random variable $E$ described by Eq.~\eqref{eq:exponential} has
\begin{align*}
    P_E(x) & = 1 - e^{-\lambda (x-x_0)},\\
    m_E(x) & = \frac{1}{\lambda}\left[1 +\lambda x_0 - (1 + \lambda x)e^{-\lambda (x-x_0)}\right],\\
    \mu_E  & = \frac{1}{\lambda}+x_0.\\
\end{align*}
Substitution in Eq.~\eqref{eq:jcont} gives
\begin{align}
    \mathcal{J}_E(x) = \frac{1+\lambda x_0-e^{-\lambda(x-x_0)}}{\lambda x + e^{-\lambda(x-x_0)}}.\label{eq:jexp}
\end{align}
Substitution in Eq.~\eqref{eq:jmean} produces
\begin{align*}
    e^{-\lambda (x-x_0)}(2 + \lambda x_0 + \lambda x) = 1 + \lambda x_0,
\end{align*}
whose solution is given by Eq.~\eqref{eq:meanexp} of the main text.

Figure~\ref{fig:sim-exp} shows the Jaccard similarity for this distribution. In this case, it is seen that the similarity decreases slowly for large $x$.

\subsection{Truncated Normal}

The truncated normal distribution of Eq.~\eqref{eq:trunnorm} has
\begin{align*}
    P_N(x) & = \frac{\erf\left(\frac{x-\mu}{\sqrt{2}\sigma}\right)+\erf\left(\frac{\mu}{\sqrt{2}\sigma}\right)}{1+\erf\left(\frac{\mu}{\sqrt{2}\sigma}\right)},\\
    m_N(x) & = \mu P_N(x) + \frac{\sigma^2}{\alpha}\left(e^{-\frac{\mu^2}{2\sigma^2}}-e^{-\frac{(x-\mu)^2}{2\sigma^2}}\right),\\
    \mu_N  & = \mu + \frac{\sigma^2}{\alpha}e^{-\frac{\mu^2}{2\sigma^2}}.\\
\end{align*}
Which, when substituted in Eq.~\eqref{eq:jcont} (omitted here) and Eq.~\eqref{eq:jmean}, gives Eq.~\eqref{eq:meannorm} of the main text.

Figure~\ref{fig:sim-norm} shows the Jaccard similarity for this distribution. As in the case of the exponential, the similarity decays slowly to the right.

\subsection{Power Law}

The random variable $P$ whose distribution is given by Eq.~\eqref{eq:pow} in the region of interest ($x > x_0$) has
\begin{align*}
    P_P(x) & = 1 - \left(\frac{x}{x_0}\right)^{-\gamma + 1},\\
    m_P(x) & = \frac{\gamma-1}{\gamma-2}x_0\left[1-\left(\frac{x}{x_0}\right)^{-\gamma+2}\right],\\
    \mu_P  & = \frac{\gamma-1}{\gamma-2}x_0.\\
\end{align*}
Substitution in Eq.~\eqref{eq:jcont} gives
\begin{align}
    \mathcal{J}_P(x) = \frac{(\gamma-1)(x/x_0)^{\gamma-2}-1}{(\gamma-2)(x/x_0)^{\gamma-1}+1}\label{eq:jaccpow}
\end{align}
and in Eq.~\eqref{eq:jmean}, gives Eq.~\eqref{eq:meanpow} of the main text.

Figure~\ref{fig:sim-pow} shows the Jaccard similarity for this distribution. Larger values of $\gamma$, which imply a faster decay of the probability with $x$, have high similarity values more concentrated around the Jaccard mean, but smaller values of $\gamma$ show a very slow decay of the similarity with growing $x$.

\section{Translation}\label{app:trans}

\begin{figure}
\centering
\begin{subfigure}{0.4\textwidth}
    \includegraphics[width=\textwidth]{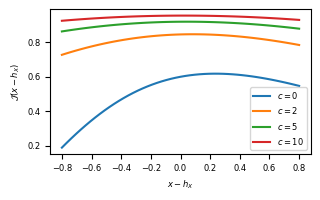}
    \caption{}
    \label{fig:translate-uni}
\end{subfigure}
\begin{subfigure}{0.4\textwidth}
    \includegraphics[width=\textwidth]{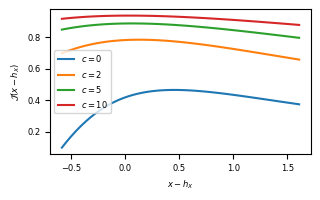}
    \caption{}
    \label{fig:translate-exp}
\end{subfigure}
\begin{subfigure}{0.4\textwidth}
    \includegraphics[width=\textwidth]{translate-norm.png}
    \caption{}
    \label{fig:translate-normb}
\end{subfigure}
\begin{subfigure}{0.4\textwidth}
    \includegraphics[width=\textwidth]{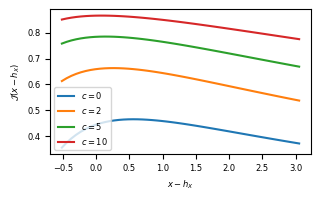}
    \caption{}
    \label{fig:translate-pow}
\end{subfigure}
    \caption{Effects of different translations on the Jaccard similarity. In each graph the similarity is plotted in the region from the percentile 10\% to the percentile 90\% of the distribution. The plots are shifted such that the medians $h_X$ are at zero. The original distributions before translation are: (a)~A uniform distribution with $\mu=1$ and $d=1$; (b)~an exponential distribution with $\lambda=1$ and $x_0=0$; (c)~a truncated normal distribution with $\mu=10$ and $\sigma=1$ (a larger $\mu$, and we use proportionally larger values of $c$ to get the same $\mu_X/c$ ratios, are used here to avoid interference from the truncation); and a power law distribution with $\gamma=2.5$ and $x_0=1$.}
    \label{fig:translateex}
 \end{figure}

In Fig.~\ref{fig:translateex} the effect of translation in the distributions used here as examples is demonstrated. The similarity profile of the original distribution (case $c=0$ in the figure) is compared with the similarity profile of the same distribution but with values shifted by $c=2, 5, 10$. For comparison, the similarity profiles of the shifted versions are shifted back by subtracting from $x$ the distribution median $h_X$. It can be seen that the Jaccard similarity is not symmetric, even for symmetric distributions. But shifting results, together with an increase in similarity values, in a more symmetric profile in the region of interest, which can be an interesting property for some applications. These results are as expected by analyzing Eq.~\eqref{eq:jtrans}.

\end{appendices}

\end{document}